\documentstyle[aps,prl,epsfig,multicol]{revtex}
\newsavebox{\LSIM}
\sbox{\LSIM}{\raisebox{-1ex}{$\ \stackrel{\textstyle<}{\sim}\ $}}
\newcommand{\lsim}{\usebox{\LSIM}}

\begin{document}

\title{Limitations of light delay and storage times in EIT experiments with condensates}
\author{D.C.~Roberts, T.~Gasenzer and K.~Burnett}
\address{Clarendon Laboratory, Department of Physics, University of Oxford,
Oxford OX1~3PU, United Kingdom}
\date{11 January 2002}
\maketitle

\begin{abstract}
We investigate the limitations arising from atomic collisions on the storage and delay times of probe pulses in EIT experiments.  We find that the atomic collisions can be described by an effective decay rate that limits storage and delay times.  We calculate the momentum and temperature dependence of the decay rate and find that it is necessary to excite atoms at a particular momentum depending on temperature and spacing of the energy levels involved in order to minimize the decoherence effects of atomic collisions.  
\\[3pt]
\end{abstract}

\begin{multicols}{2}
Electromagnetically induced transparency (EIT) \cite{EIT} allows a medium that is opaque to be made transparent around its resonant frequency. 
Recent experiments using EIT of cold trapped atomic gases have shown that one can delay light pulses by reducing the group velocity to speeds as low as $1\,$m/s \cite{haudelay,Inouye00}. Moreover it is possible to store them for up to $1\,$ms before regenerating them \cite{hauhalt} (for a recent review see \cite{lukin}).  
There are two main factors that affect the photon delay or storage time in EIT experiments \cite{lukin,fleish2}.  
Firstly, the delay time is limited by the finite frequency width of this transparency window.  We shall assume that the pulse duration is sufficiently long that its bandwidth fits within the transparency window.  
Secondly, the decay of the dark state arising from atomic collisions, as well as from possible couplings to states outside the system, limits the delay and storage times in these experiments.   
In this paper we discuss how the decay of the dark state is related to the delay or storage times. 
We study the dependence of the decay width introduced by collisional atomic interactions on temperature and on the momentum transferred from the photons to the atoms.

\begin{figure}[ht]
\begin{center}
\epsfig{file={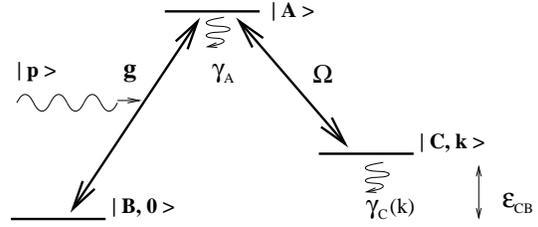},height=3cm,width=7 cm,angle=0}\\[5mm]

\caption{\label{raman} We consider 3-level atoms illuminated by a single quantized (probe) mode with coupling $g$ and a classical laser field characterized by the Rabi frequency $\Omega$.  We assume the condensate atoms to be in a coherent state $|B,0\rangle$. The lasers transfer the atoms to the internal level $C$, into the translational state $|C,k\rangle$ with defined momentum $k$. $\gamma_A$ and $\gamma_C(k)$ characterize the decay rates from $|A \rangle $ and $|C,k \rangle$, respectively.}
\end{center}
\end{figure}

We consider a homogeneous gas of atoms whose internal structure may be described as a three-level system as shown in Figure \ref{raman}.
Homogeneity should be a good approximation for the central region of the large alkali-metal condensates achieved in present day experiments.
The ground state $|B,0 \rangle$, which describes a coherently populated, weakly interacting Bose-Einstein condensate with a mean number of $N_0$ atoms, is resonantly coupled to the excited state $|A \rangle$ by a weak probe pulse. 
This probe pulse is described by a single-mode quantized electromagnetic field with frequency $\omega_p$. 
Its dipolar coupling strength to the atoms is $g$. 
For convenience, we take the probe pulse to be initially in a number state $|p;n\rangle$ with $n$ photons.
The metastable translational state $|C,k \rangle$ of atoms with c.m.~momentum $k$ in internal state $|C\rangle$ is resonantly coupled to $|A \rangle$ with a classical laser field with frequency $\omega_c$ characterized by the Rabi frequency $\Omega$.   
We describe the decay of the excited state $|A \rangle$ due to spontaneous emission with a rate $\gamma_A$. 
\vspace*{-3mm}
%=============================================================================
%=============================================================================
\vspace*{-3mm}
The population of the state $|C,k \rangle$ decays predominantly due to 2-body elastic collisions with the atoms in internal state $|B\rangle$ which conserve the internal atomic quantum number.
This decay may be described by a rate $\gamma_C(k)$ depending on the momentum of the decaying mode if the modes with momenta different from $k$ as well as the condensate of atoms in internal level $|B\rangle$ are close to thermal equilibrium. $\gamma_C(k)$ may be calculated perturbatively using a linear response approach to the mode operator time evolution \cite{Gasenzer02}.

The dynamics of the system we consider is described by the following system of equations of motion for the atomic operators $\hat A$, ${\hat C}_k$ and the probe photon operator $\hat a$:
%
%\begin{equation}
%\label{eq1}
%  \frac{d}{dt}
%  \left(\begin{array}{c} \hat A \\ \hat C \\ \hat a \end{array}\right)
%  = -i\,\left(\begin{array}{ccc}
%    0 & \Omega & g\sqrt{N_B} \\ 
%    \Omega & -i\gamma_C(k)/2 & 0 \\ 
%    g\sqrt{N_B} & 0 & 0 \end{array}\right)
%  \left(\begin{array}{c} \hat A \\ \hat C \\ \hat a \end{array}\right)
%  -\left(\begin{array}{c} 0 \\ {\hat F}_k \\ 0 \end{array}\right).    
%\end{equation}
%
%
\begin{eqnarray}
\label{eq1}
  \dot{\hat A} 
  &=& -i\Omega\,\hat C_k -i g\sqrt{N_0}\,\hat a, \nonumber\\
  \dot{\hat C_k}
  &=& -i\Omega\,\hat A-(\gamma_C(k)/2)\,\hat C_k -{\hat F}_{C,k}, \nonumber\\
  \dot{\hat a} 
  &=& -ig\sqrt{N_0}\,\hat A-(\gamma_A/2)\,\hat a -{\hat F}_A.
\end{eqnarray}
Here we have chosen an interaction picture where the fast evolution due to the energy splitting of the states has been factored out of the operators, and the resonance conditions $\epsilon_{AB}-\epsilon_{CB}-\tilde{\epsilon}_k^C=\hbar\omega_c$, $\epsilon_{AB}=\hbar\omega_p$ have been taken into account. 
Here, $\tilde{\epsilon}_k^C=\hbar^2 k^2/(2m)$ is the c.m.~kinetic energy of an atom in mode $k$ and internal state $|C\rangle$.  
${\hat F}_{C,k}$ and ${\hat F}_A$ are the Langevin noise operators corresponding to the decay of the operators $\hat C_k$ and $\hat a$ respectively which provide for the mode operators to obey Bose commutation relations and the correlation functions of these operators to decay to their equilibrium values \cite{MandelWolf,Gasenzer02}.

The recently achieved light propagation slowing and storage of light pulses in cold atomic gases rely on the possibility to create a dark state.
A dark state $|D(n,\theta) \rangle$ which is consistent with our assumption of a probe beam being initially in a number state and a condensate in a coherent state $|B,0\rangle$ has the form of a ``beam-splitter state'':
\begin{eqnarray}
\label{dark}
  |D(n,\theta)\rangle 
  &=& (\cos \theta)^{n}\sum_{m=0}^n \sqrt{\frac{n!}{m!(n-m)!}}
    (-\tan\theta)^m
  \nonumber\\
  &&\ \times |C,k;m\rangle|p;n-m\rangle
\end{eqnarray}
where $|C,k;m\rangle$ is the number state where $m$ atoms are in mode $k$ in level $C$.
If the mixing angle is defined through
\begin{equation}
  \tan\theta = \frac{g \sqrt{N_0}}{\Omega},
\end{equation} 
and provided one neglects $\gamma_C(k)$, the state (\ref{dark}) does not change in time except for an overall phase. 
By adiabatically adjusting $\Omega$ and thereby varying $\theta$ from $0$ to $\pi/2$, one can transfer the initial photon population into collective atomic excitations in $|C,k \rangle$.  
The reverse can be achieved by adiabatically changing $\theta$ back to $0$. 
In \cite{fleish1,fleish2} this is described in terms of dark-state polaritons. 
In this idealized process  the sum of photons in the probe beam state and atoms in $|C,k \rangle$, i.e.~$n_p(t)+n_{C,k}(t)=\langle{\hat a}^\dagger(t)\hat a(t)\rangle+\langle{\hat C}_k^\dagger(t){\hat C}_k(t)\rangle$, remains constant and the model describes the stopping and regenerating of a light pulse in an atomic medium.  

This simple picture no longer fully holds once decay is taken into account.  
We have studied the effects of $\gamma_C(k)$ by numerically calculating $n_p(t)+n_{C,k}(t)$.
To this end we use the Heisenberg equations of motion (\ref{eq1}) with respect to an initial dark state (\ref{dark}), assuming the corresponding energies to have small imaginary components as given in (\ref{eq1}).   
Note that we have in  principle to take into consideration the corresponding Langevin noise operators ${\hat F}_{C,k}$, ${\hat F}_A$ when calculating the time evolution of the number operators.
One may show, however, that a quantum regression theorem holds for the correlation functions of two normal ordered operators \cite{Gasenzer02}.
Therefore (\ref{eq1}) may be used, neglecting the Langevin terms, for the calculation of their time evolution. 

\begin{figure}[ht]
\begin{center}
\epsfig{file={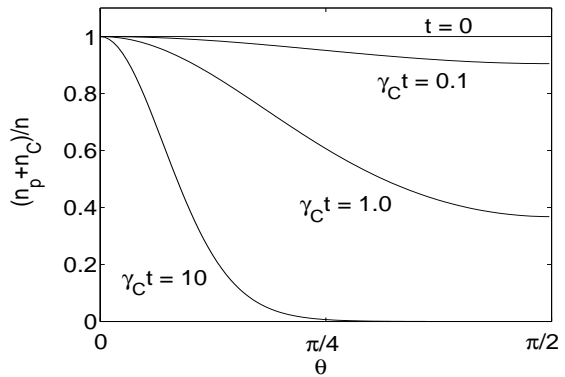},height=5 cm,width= 7.5 cm,angle=0}\\[3mm]
\caption{\label{theta}  Number of photons in the weak probe beam plus the number of atomic excitations in $|C,k \rangle$, $n_p+n_{C,k}$, as a function of $\theta$ for different times using experimental parameters \protect\cite{haudelay}: $\Omega = (2\pi)5.61\,$MHz, $\gamma_A=(2\pi)10\,$MHz. We also assume $\gamma_C(k)<g\sqrt{N_0}$. ($N_0\simeq10^6$ atoms, $n\simeq 3\cdot10^4$ photons.)}
\end{center}
\end{figure}

We see in Figure \ref{theta} that $n_p+n_{C,k}$ decays faster the closer $\theta$ is to $\pi/2$, or equivalently, $\Omega$ is to zero.  
This faster decay arises because more photons are converted into atomic excitations.  
Apart from this, the decay is dominated by $\gamma_{C}(k)$ and as long as this is reasonably small (i.e.~$\gamma_{C}(k)<g \sqrt{N_0}$, which is well satisfied in present experiments, e.g.~$\gamma_{C}(k)<(2\pi)10\,$kHz, $g \sqrt{N_0}\simeq(2\pi)15\,$MHz in \cite{hauhalt}), the populations $n_p+n_{C,k}$, $n_p$ and $n_{C,k}$ all decay at the same rate.
%=============================================================================
%=============================================================================
In the light storage experiments, the system remains at $\theta=\pi/2$ (i.e.~$\Omega=0$) for much longer than it spends in transition.  
For example, in \cite{hauhalt} the delay time plus the time it took to turn on and off the coupling laser was on the order of microseconds whereas the maximum storage time was on the order of milliseconds.  
Since the rate of decay is highest when the photons are stored ($\theta=\pi/2$) and the storage time far exceeds the delay and turn-on/off times, we approximate that all of the decay of $n_p+n_{C,k}$ occurs while the photons are stored as atomic excitations.  
The maximum storage time $\tau_s$ is then given by $\tau_s\lsim 1/\gamma_C(k)$.

For slow light experiments, the characteristic decay time, $\tau_d$, at which $n_p+n_{C,k}$ has decayed to a fraction $1/e$ of its initial value $n$, has to be calculated numerically.  
Assuming that effects other than collisions are irrelevant \cite{othereffects}, we estimate $\tau_d$ for coherent slow light propagation in a Bose-condensed gas at $T=0$ from the time evolution of the dark state (\ref{dark}), for the experimental parameters from \cite{haudelay}: $\Omega=(2\pi)5.61\,$MHz and $g \sqrt{N_0}\simeq(2\pi)10\,$MHz.
We find that $\tau_d \approx 1.2/\gamma_C(k)$.
As can be seen in Figure \ref{theta} $\tau_d$ is in general greater than $1/\gamma_C(k)$.

We have seen that the decay caused by collisions and characterized by $\gamma_C(k)$ limits the maximum time for the storage of light pulses in an atomic condensate.
We have calculated $\gamma_C(k)$ for the decay of a single excited mode of atoms in a different internal state than the condensate atoms, using stationary perturbation theory in the interactions beyond the quadratic Hartree-Fock-Bogoliubov hamiltonian. 
The results of this calculation can also be derived using a time-dependent linear response approach \cite{Gasenzer02}. 
In the following we study the dependence of $\gamma_C(k)$, and hence of $\tau_d$ and $\tau_s$, on momentum $k$ and temperature. 
We consider a homogeneous dilute condensate of $N_0$ particles dominated by two-body collisions.
Their strength is determined by the s-wave scattering length $a$, which we assume to be the same for collisions either of atoms within level $B$ or between atoms in $B$ and $C$.  
An extension of our methods to the case of trapped condensates is straightforward.
We assume the condensate is confined in a box of volume $V$ and apply periodic boundary conditions. 
Collisions between atoms in $|B\rangle$ change the free particle dispersion relation to the Bogoliubov quasi-particle dispersion relation 
\begin{equation}
  \tilde{\epsilon}^B_k=\mu\,y_k \sqrt{2+y_k^2}
\end{equation}
where $\mu=(\hbar k_0)^2/2m$ and $y_k=k/k_0$. The inverse of the healing length, $k_0=(8 \pi n_0 a)^{1/2}$, separates the phonon-type from the free-particle type region of the quasi-particle spectrum. 
To the leading order, only the spin-conserving collisions between atoms in $|B \rangle$ and $|C\rangle$ affect the excitations in $|C,k \rangle$,  
causing a non-zero $\gamma_C(k)$. 
Collisions involving atoms in $|A\rangle$ can be ignored because of its negligible population.  
Finally, collisions between atoms in internal state $|C\rangle$ can be neglected as there is no macroscopically occupied c.m.~mode.
We neglect the energy shift of the levels due to interactions between excitations in $|C\rangle$ and $|B\rangle$, which arises at the same order as  $\gamma_C(k)$ \cite{Gasenzer02}. 

The collisional decay rate can be separated into the Beliaev-like and Landau-like decay rates (hereafter referred to as Beliaev and Landau decay rates), i.e. $\gamma_C(k)=\gamma^{\mathrm B}_C(k)+\gamma^{\mathrm L}_C(k)$.
The Beliaev decay rate, corresponding to a particle in $|C,k \rangle$ colliding with a condensate atom in $|B,0\rangle$, producing one quasiparticle in $|B,i \rangle$  and one particle excitation in $|C,j\rangle$, both of lower energy, $z_k^C=z_i^C+z_j^B$ ($z_{j}^B=y_{j}\sqrt{2+y_{j}^2}$, $z_{k}^C=y_{k}^2$), is given by
\begin{eqnarray}
\label{beliaev}
  \gamma^{\mathrm B}_C(k)
  &=& \frac{a \mu k_0}{\hbar y_k}\int_0^{z^C_k} dz^B_j 
      \left(1-\frac{1}{\sqrt{1+(z^B_j)^2}} \right)
  \nonumber\\
  &&\qquad\qquad \times\ (1+n^C_i+n_j^B).
\end{eqnarray}
with populations $n_{j}^B=(\exp[z^B_{j}\mu/k_BT]-1)^{-1}$ and $n_{i}^C=(\exp[(z_{k}^C+z_{CB}-z^B_j)\mu/k_{B}T]-1)^{-1}$. ($z_{CB}=\epsilon_{CB}/\mu$).
The Landau decay rate is caused by the reverse processes where a particle in $|C,k \rangle$ collides with a quasiparticle in $|B,j \rangle$ to produce a particle excitation in $|C,j\rangle$ with higher energy $z_i^C=z_k^C+z^B_j$. It is given by
\begin{eqnarray}
\label{landau}
  \gamma^{\mathrm L}_C(k)
  &=& \frac{a \mu k_0}{\hbar y_k}\int_0^\infty dz^B_j 
      \left(1-\frac{1}{\sqrt{1+(z^B_j)^2}} \right)
  \nonumber\\
  &&\qquad\qquad \times\ (n_j^B-n_i^C),
\end{eqnarray}
where $n_{i}^C=(\exp[(z_{k}^C+z_{CB}+z^B_j)\mu/k_B T]-1)^{-1}$.  
Whereas Beliaev processes are restricted to modes whose energies are less than that of decaying particles, Landau processes can involve any mode with an appreciable population.  
This causes the Landau decay to dominate at higher temperatures.
At $T=0$ only the Beliaev rate survives giving
\begin{equation}
\label{decay0}
  \gamma_C(k)_{T=0}
  = a\omega_0 k_0 y_k \left(1-\frac{\ln(y_k^2+\sqrt{1+y_k^4})}{y_k^2}\right).
\end{equation} 
At high momentum where $k\gg k_0$, $\gamma_C(k)_{T=0} \approx \gamma_{KT}(k)$, where  $\gamma_{KT}(k)=n_0 \sigma v$ is the collision rate from classical kinetic theory, assuming the particles have speed $v=\hbar k/m$ and cross sectional area $\sigma = 4 \pi a^2$. At low momentum $k\ll k_0$, we obtain
$\gamma_C(k)_{T=0} \approx \hbar k^5/96 m \pi n_0$.
This low-momentum limit of the decay rate has the same $k^5$ dependence as the $T=0$ Beliaev decay rate for quasi-particle excitations in a single state condensate \cite{beliaev,giorgini}, from which it only differs by a constant numerical factor. 

\begin{figure}[ht]
\begin{center}
\epsfig{file={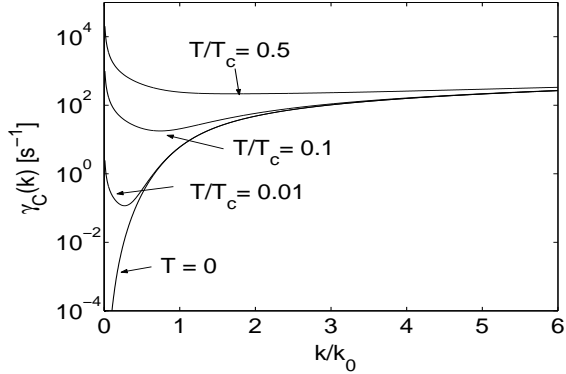},height=5 cm,width= 7.5 cm,angle=0}\\[3mm]
\caption{\label{decay} The decay rate $\gamma_C(k)$ due to atomic collisions, as a function of normalized momentum $k/k_0$ for different temperatures, using the experimental parameters \protect\cite{haudelay}: 
%$1.2\cdot10^{6}$ 
$^{23}$Na atoms, $a=2.8\,$nm, $T_c=435\,$nK, $n_0=8\cdot10^{13}\,$cm$^{-3}$, $\mu/\hbar=(2\pi)\,1.21\,$kHz and $\epsilon_{CB}/\hbar=(2\pi) 1.8\,$GHz.}
\end{center}
\end{figure}

\begin{figure}[t]
\begin{center}
\epsfig{file={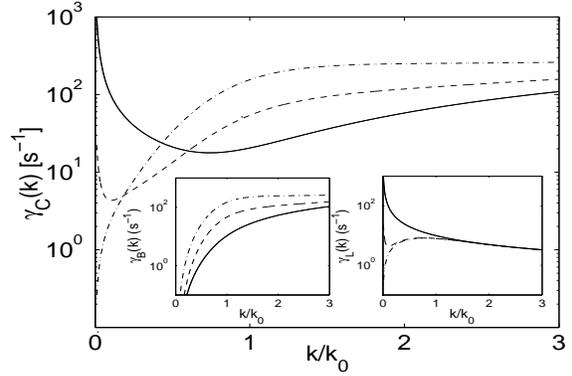},height=5 cm,width= 7.5 cm,angle=0}\\[3mm]
\caption{\label{zcb} The decay rate $\gamma_C(k)$, as a function of $k/k_0$ for different values of $z_{CB}=\epsilon_{CB}/\mu$,  for the same experimental parameters as used in Figure \ref{decay} and $T/T_c=0.1$.  We chose $z_{CB}=0.0001$ (dash-dotted line), $z_{CB}=0.01$ (dashed line), and $z_{CB}>10$, where $n^C_i \approx 0$ (solid line).  The two insets show $\gamma^{\mathrm B}_C(k)$ and $\gamma^{\mathrm L}_C(k)$ for the same parameters. }
\end{center}
\end{figure}

In Figure \ref{decay} we plot $\gamma_C(k)$ as a function of $k/k_0$ for different temperatures using experimental parameters of \cite{haudelay}.  
As the temperature increases from $T=0$, Landau processes start to dominate at low $k$ while Beliaev processes remain dominant at high $k$, creating a minimum decay rate at a finite $k$.  
Eventually the minimum disappears because at that point Landau processes dominate at all $k$.   
The minimum of $\gamma_C(k)$ at a finite $k$ develops because of the separation $\epsilon_{CB}$ of the energy levels $|B,0\rangle$ and $|C,0 \rangle$.
Note that for the condensate in \cite{haudelay} with a length of $230\,\mu$m the minimum $y_k=k/k_0$ is of the order of $5\cdot10^{-3}$, such that the description of the loss from $|C,k\rangle$ by a decay rate is expected to be justified.  
%=============================================================================
%=============================================================================
To emphasize the importance of $\epsilon_{CB}$,  we plot $\gamma^{\mathrm B}_C(k)$, $\gamma^{\mathrm L}_C(k)$, and $\gamma_C(k)$ as functions of $k$ for different values of $\epsilon_{CB}$ in Figure \ref{zcb}.  
We see that, as $\epsilon_{CB}$ is increased, the Landau rate significantly increases at low $k$ because a large $\epsilon_{CB}$ causes $(n^B_j-n^C_i)$ in eqn.~(\ref{landau}) to approach $n^B_j$ while $(n^B_j-n^C_i)$ approaches zero as $k \rightarrow 0$ for $\epsilon_{CB} \rightarrow 0$.  
Furthermore, the Beliaev rate decreases slightly as $\epsilon_{CB}$ increases because $n^C_i \approx 0$ in eqn.~(\ref{beliaev}) at large $\epsilon_{CB}$.  
We can see that the minimum in $\gamma_{C}$ develops for sufficiently large $\epsilon_{CB}$ at certain temperatures because the Landau rate dominates at low $k$ while the Beliaev rate dominates at high $k$.
As the temperature increases, the momentum transfer required to minimize $\gamma_C(k)$ also increases.  For example, using $T/T_c=0.5$ and the experimental parameters used in Figure \ref{decay}, we find $\tau_s=0.5\,$ms at $y_k=0.1$ and $\tau_s=4.1\,$ms at $y_k=1$.  

We have shown in this paper that the storage time for light pulses in atomic condensates using dark-state based EIT schemes is limited by loss caused by collisions between particles and quasiparticles in the atomic ensemble.
We found that one can minimize this loss in a homogeneous BEC by adjusting the momentum $k$ transferred to the atoms. 
The optimum $k$ depends on the characteristics of the three-level system and the temperature.
In this respect the collisional decay in a system where the atoms in the excited translational mode are in a different internal state than the condensate atoms shows an important property distinct from the case with only one internal state.
%=============================================================================
%=============================================================================

This work was financially supported by the British Marshall Foundation (D.C.R.), the A.~von Humboldt-Foundation, the European Community under contract no.~HPMF-CT-1999-0023 (T.G.), and the United Kingdom EPSRC.

\end{multicols}


\begin{thebibliography}{9}
\bibitem{othereffects}
The coupling laser may induce also non-resonant transitions to states outside the three-level system \cite{haudelay}.
\end{thebibliography}
\end{document}